\documentclass[12pt,a4paper]{article}
\usepackage[utf8]{inputenc}
\usepackage{amsmath}
\usepackage{amsfonts}
\usepackage{amssymb}
\usepackage{hyperref}
\usepackage{graphicx}
\usepackage[margin=1in]{geometry}

\title{\huge{\textbf{Nonparaxial accelerating \\ Bessel-like beams}}}
\author{Ioannis D. Chremmos and Nikolaos K. Efremidis\\\\
\textit{Department of Applied Mathematics, University of Crete}\\
\textit{Heraklion 71409, Greece}\\
\textit{jochremm@central.ntua.gr, nefrem@tem.uoc.gr}}
\date{}

\begin{document}
\setlength{\abovedisplayskip}{20pt}
\setlength{\belowdisplayskip}{20pt}
\setlength{\abovedisplayshortskip}{20pt}
\setlength{\belowdisplayshortskip}{20pt}
\maketitle

\begin{abstract}
A new class of nonparaxial accelerating optical waves is introduced. These are beams with a Bessel-like profile that are capable of shifting laterally along fairly arbitrary trajectories as the wave propagates in free space. The concept expands on our previous proposal of paraxial accelerating Bessel-like beams to include beams with subwavelength lobes and/or large trajectory angles. Such waves are produced when the phase at the input plane is engineered so that the interfering ray cones are made to focus along the prespecified path. When the angle of these cones is fixed, the beams possess a diffraction-free Bessel profile on planes that stay normal to their trajectory, which can be considered as a generalized definition of diffractionless propagation in the nonparaxial regime. The analytical procedure leading to these results is based on a ray optics interpretation of Rayleigh-Sommerfeld diffraction and is presented in detail. The evolution of the proposed waves is demonstrated through a series of numerical examples and a variety of trajectories.
\end{abstract}

\section{Introduction}

Spatially accelerating optical beams are currently very popular forms of structured light. The interest in them was triggered in 2007 when the Airy wavepacket ---previously known within quantum mechanics \cite{Berry1979}--- was introduced into the optical domain \cite{Siviloglou2007, sivil-prl2007}. Such beams have the two salient properties of being diffraction-free (or diffraction-resisting in their actual, finite-energy implementations) in addition to shifting their intensity features along parabolic trajectories in free space. Soon it was found that these two properties define a general class of paraxial waves \cite{Bandres2009}, with Airy beams being the most straightforward to produce due to their experimentally simple to generate Fourier transform \cite{sivil-prl2007}. Moreover, it was proved that the parabolic trajectory is unique in this class of waves, in the sense that diffraction-free paraxial waves can accelerate only along parabolas \cite{Bandres2009, Unnikrishnan1996}. Besides challenging our everyday experience of light beams that propagate along straight paths, accelerating beams have found a diversity of useful applications including optical manipulation \cite{Christodoulides_2008_Riding, Baumgart_2008, Zhang_2011_Morphing}, curved plasma filaments \cite{Polynkin_2009}, laser microfabrication \cite{Mathis_2012}, Airy plasmons \cite{Salandrino_2010} and abruptly autofocusing waves \cite{Efremidis2010}. For a comprehensive review of the field the reader is referred to \cite{Hu2012Springer} and to the references therein.

The acceleration of an optical beam in space is the result of a particular interference pattern, which ray optics explain through the concept of caustics \cite{Nye1999}. Using this principle, accelerating beams with arbitrary convex trajectories \cite{Greenfield2011, Chremmos_2011_AAF, Froehly_2011} (hence strictly not diffraction-free) have been obtained in 1D Cartesian coordinates \cite{Greenfield2011, Froehly_2011} and in cylindrical coordinates \cite{Chremmos_2011_AAF} in the form of caustic surfaces under revolution. Interestingly enough, in these two cases both the physics and the mathematics of caustics are identical due to the fact that the ray (a single line) can not distinguish a difference when it propagates in a Cartesian of in a radial coordinate. Engineering the spectrum of the beam on the Fourier plane is an efficient method for generating such waves \cite{Chremmos2012_PRA, Chremmos_2011_Fourier}. As their ray structure suggests, optical beams that accelerate along curved caustics inevitably have an asymmetric power profile. Indeed, the mathematical expression of the optical field through a caustic is proportional to the Airy function, thus proving  that power is confined to the side of the caustic where the oscillations occur. On the other hand, the only other known class of diffraction-free beams, namely the propagation-invariant beams \cite{Piestun_1998}, includes waves with symmetric profiles that follow from the intereference of conical ray bundles. However such beams propagate along straight trajectories. The most famous example are Bessel beams \cite{Durnin1987} with their numerous applications \cite{Mcgloin_2005}.

In view of the above, it would be natural to ask whether there can be \textit{hybrid} beams that can accelerate while maintaining a diffraction-free (or at least diffraction-resisting) circularly symmetric profile. This question was first addressed in \cite{Chremmos_2013_Bessel}, where a systematic method was proposed to design Bessel-like beams capable of accelerating along arbitrary trajectories in free space. The method is based on an appropriate modification of the conical ray pattern of standard Bessel beams to create deformed ray cones that converge along a pre-specified focal line. The modified rays emerge when an phaseless initial wavefront is modulated by an appropriate phase mask that is calculated numerically by the procedure detailed in \cite{Chremmos_2013_Bessel}. The proposed beams were observed experimentally in \cite{Zhao_2013}.

Interestingly, and despite the popularity of accelerating waves over the last 5 years, there have been only two previous efforts to create accelerating Bessel-like beams. In \cite{Jarutis_2009} spiraling Bessel-like beams were obtained through the combination of the standard axicon lens with a holographic spiral phase pattern. Experimental demonstrations of this method were reported in \cite{Matijosius_2010}. The same principle was used in \cite{Morris_2010}, where the phases of two counter-rotating spiraling beams were superposed to produce snaking Bessel-like beams capable of propagating around obstructions. However, these works were restricted to the two mentioned trajectories and no systematic method was devised for extending to arbitrary trajectories. It should also be noted that the idea of snaking beams had appeared many years before in \cite{Rosen_1995}, where three "sword beams", produced by distinct phase masks, were cascaded in the $z$ direction to create a zigzag focal line. Such piecewise-defined beams are however not accelerating. It is also important not to confuse the accelerating Bessel-like beams introduced in \cite{Chremmos_2013_Bessel} with the so-called helicon \cite{Paterson_1996_Helicon} or helico-conical beams \cite{Alonzo_2005_Helico}. Helicon beams result from the superposition of two standard Bessel beams and fall under the category of generalized propagation-invariant waves whose profile is steady in a rotating frame \cite{Piestun_1998}. Helico-conical beams on the other hand have a transverse profile that resembles a dilating arithmetic (Archimedean) spiral and hence are neither diffraction-free nor accelerating.

In present work we take the idea of \cite{Chremmos_2013_Bessel} a step further and explore Bessel-like accelerating beams in the nonparaxial regime. Nonparaxial propagation occurs when the oscillations of the beam's profile in the transverse direction are in a scale comparable to or faster than the wavelength. Under such conditions, the wave dynamics obey the vector Helmholtz equation $(\nabla ^2 + k^2) \bf E = \bf 0 $ and the corresponding dispersion relation $k_z = (k^2 - k_x^2 - k_y^2)^{1/2}$ ($k$ being the wave number), which should be contrasted to the scalar paraxial wave equation $2iku_z + u_{xx}=0$ and the corresponding quadratic dispersion law $k_z = k - (k_x^2 + k_y^2)/2k$.

The interest in nonparaxial accelerating waves has recently increased as researchers try to enhance the potentiality of accelerating beams for applications by bending light beyond paraxial angles. The nonparaxial regime of Airy beams has already been investigated in several works \cite{Novitsky_2009, Carretero_2009, Torre_2010, Kaganovsky_2012}, revealing a significantly different ray and caustic structure as well as the effect of low-pass spectral filtering due to the evanescent propagation of wave numbers $k_x^2 + k_y^2 > k^2$. The first effort to design arbitrary 2D and 3D micron-scale caustics by applying a phase mask to the input wavefront was reported in \cite{Froehly_2011}. Contrary to the paraxial case, the phase function required for nonparaxial caustics can rarely be determined in closed form. Collapsing caustic surfaces of revolution were also produced in the same work following the idea of pre-engineered abruptly autofocusing beams \cite{Efremidis2010, Chremmos_2011_AAF}. The same group later reported the propagation of femtosecond pulses along arbitrary convex caustics and proved that circular trajectories are diffraction-resisting in the sense of preserving the intensity FWHM of the beam's main lobe in the direction normal to the caustic \cite{Courvoisier_2012_Femto}. Such pulses were subsequently used to create curved ablation trenches in diamond and silicon samples \cite{Mathis_2012}. In a different approach, nonparaxial beams with circular trajectories were produced by taking the angular spectrum of high-order Bessel wavefunctions and isolating the forward-propagating components  \cite{Kaminer_2012}. This idea was also adopted in \cite{Zhang_2012_Nonparaxial} where nonlinear propagation of the produced beams was additionally investigated. Finally, general families of 2D diffraction-resisting nonparaxial beams have been derived from separable solutions of the Helmoltz equation in elliptic and parabolic coordinates, termed Mathieu and Weber beams, respectively \cite{Zhang_2012_Weber, Aleahmad_2012_PRL}. The former write elliptic caustics and generalize the previously reported circular-caustic beams, while the latter write parabolic caustics and behave as nonparaxial counterparts of standard Airy beams. More complicated families of fully vectorial 3D beams have also been retrieved from separable solutions of the Helmholtz equation in spherical and spheroidal coordinates \cite{Aleahmad_2012_PRL}.

In the following sections we introduce nonparaxial accelerating Bessel-like beams and present in detail the analysis leading to their design. Making use of the mathematical tool of vector potentials, the problem is reduced to studying the dynamics of a scalar wavefunction according to the Rayleigh-Sommerfeld formula of diffraction. We find that the procedure of \cite{Chremmos_2013_Bessel} can indeed be generalized in the nonparaxial domain and leads to beams with a diffraction-resisting Bessel-like profile owed to the modified conical superposition. Still, there is a new interesting feature that sets the nonparaxial case clearly apart: the symmetric Bessel-like profile forms on a plane that is normal to the focal trajectory. This further suggests that ---in contrast to the paraxial case \cite{Bandres2009}--- nonparaxial diffraction-free (or diffraction-resisting) accelerating waves should be sought as propagation-invariant waves with a dynamically changing plane of observation that stays normal to the trajectory of the beam. This property has been evidenced in the particular case of 2D beams that maintain a constant lobe width in the direction normal to their circular caustic \cite{Courvoisier_2012_Femto}. However, this property for arbitrarily accelerating non-caustic 3D beams is reported here for the first time to our knowledge.

\section{Analysis}

\subsection{Vector beam formulation}

In paraxial optics, the EM beams are silently assumed to have a single transverse electric field component, e.g. ${\bf E} = \hat{\bf X}{E_X}(X,Y,Z)$, which is of course a violation of Gauss law $\nabla \cdot {\bf E} = 0$. This assumption is acceptable only when the beam's width is much smaller than the diffraction length \cite{Lax_1975}. In the nonparaxial regime, however, the validity of this approximation is questionable and one is required to take into account the complete vector structure of the EM waves.

The polarization of EM waves can be handled conveniently with the aid of auxiliary vector potentials of the magnetic $(\bf A)$ or electric $(\bf F)$ type \cite{Balanis_2012}. To analyze our nonparaxial beams and without loss of generality, we here consider $\text{TE}^Y$ waves, i.e. waves with zero electric component in the $Y$ direction. Such waves are derived from an electric vector potential with a single $y$ component ${\bf F} = \hat{\bf Y} u(X,Y,Z)$ through 
\begin{eqnarray}
{\bf{E}} &=&  - \nabla  \times {\bf{F}} \nonumber \\
{\bf{H}} &=& i\omega {\varepsilon}{\bf{F}} + \frac{i}{{\omega {\mu}}}\nabla \left( {\nabla  \cdot {\bf{F}}} \right)\
\label{Eq:Potential}
\end{eqnarray}
where the scalar function $u$ satisfies the homogeneous Helmholtz equation  $\left( {{\nabla ^2} + {k^2}} \right)u = 0$, $k=\omega (\epsilon \mu)^{1/2}$ being the free space wavenumber. This formulation allows us to focus our attention on the diffraction dynamics of scalar function $u$ and subsequently use \eqref{Eq:Potential} to obtain the EM components. To gain some further insight, such a beam will have a ``strong" transverse electric field component $E_X = u_Z$ \footnote[1]{Throughout the paper, the $X$ and $Y$ (or $x$ and $y$) subscripts indicate partial derivatives. The same subscripts indicate vector components \textit{only} when used with the electric and magnetic field, e.g. in $E_X$, $H_Y$ etc.} and a ``weak" longitudinal electric field component $E_Z =  - u_X$ that should get stronger the more we depart from the paraxial regime. On the other hand the magnetic field has a ``strong" $i\omega \mu {H_Y} = {u_{XX}} + {u_{ZZ}}$ (due to the $u_{ZZ}$  term) a ``weak" $i\omega \mu {H_Z} =  - {u_{YZ}}$ and an even ``weaker" $i\omega \mu {H_X} =  - {u_{XY}}$. The ``strong" transverse EM components $E_X$ and $H_Y$ are responsible for the transfer of optical power in the $Z$ direction with density $\frac{1}{2}{\mathop{\rm Re}\nolimits} \left( {{E_X}H_Y^*} \right)$, while the other components combine to account for the redistribution of power in the transverse plane as the beam propagates.

\subsection{Diffraction and ray dynamics}

The diffraction of a scalar wave according to the Helmholtz equation is expressed by the Rayleigh-Sommerfeld formula \cite{GoodmanFourier}
\begin{equation}
u\left( {\bf{R}} \right) =  - \frac{k^3 Z}{{2\pi }}\iint {u\left( {\bf{r}} \right)g\left( k{\left| {{\bf{R}} - {\bf{r}}} \right|} \right){e^{ik\left| {{\bf{R}} - {\bf{r}}} \right|}}d{\bf{r}}} 
\label{Eq:Rayleigh-Sommerfeld}  
\end{equation}
where $g(r)=ir^{-2}-r^{-3}$, ${\bf R} = (X,Y,Z)$ is the point of observation and ${\bf r} = (x,y,0)$ spans the input plane. In propagation distances $Z$ of a few wavelengths, we have $k|{\bf R} - {\bf r}| \geq kZ \gg  1$  and the ${\left( k \left| {{\bf{R}} - {\bf{r}}} \right| \right) ^{ - 3}}$ term is rather weak. For the purposes of our analysis we assume that the input condition has the form of a slowly varying envelope times a phase factor:  $u({\bf r})=U({\bf r})e^{iQ({\bf r})}$. In a stationary-phase approximation of the above integral, one requires that the partial derivatives of the total phase $P({\bf R},{\bf r}) \triangleq Q({\bf r})+k|{\bf R}-{\bf r}|$ with respect to $x$ and $y$ vanish or
\begin{equation}
{\bf{R}} = {\bf{r}} + \left| {{\bf{R}} - {\bf{r}}} \right|\left( {\frac{{{Q_x}}}{k},\frac{{{Q_y}}}{k},\frac{Z}{{\left| {{\bf{R}} - {\bf{r}}} \right|}}} \right)
\label{Eq:Stationarity}
\end{equation}
where the equation for the $Z$-coordinates is obvious. The stationary-phase approach provides a ray-optics picture of the diffraction phenomenon, whereby Eqs. \eqref{Eq:Stationarity} describe a light ray from point $\bf r$ to point  $\bf R$ with the length of the ray $|{\bf R} - {\bf r}|$  acting as a parameter. This ray has a wave vector $k({\bf R}-{\bf r})/|{\bf R}-{\bf r}|$  or equivalently ${\bf{k}}({\bf r}) = \left( {{Q_x},{Q_y},{{\left( {{k^2} - Q_x^2 - Q_y^2} \right)}^{1/2}}} \right)$  and contributes at $\bf R$ a wave with phase $Q + {\bf{k}} \cdot ({\bf R}-{\bf r}) = P({\bf R},{\bf r})$.

\begin{figure}[tb]
\centering
\includegraphics[width=0.7\textwidth]{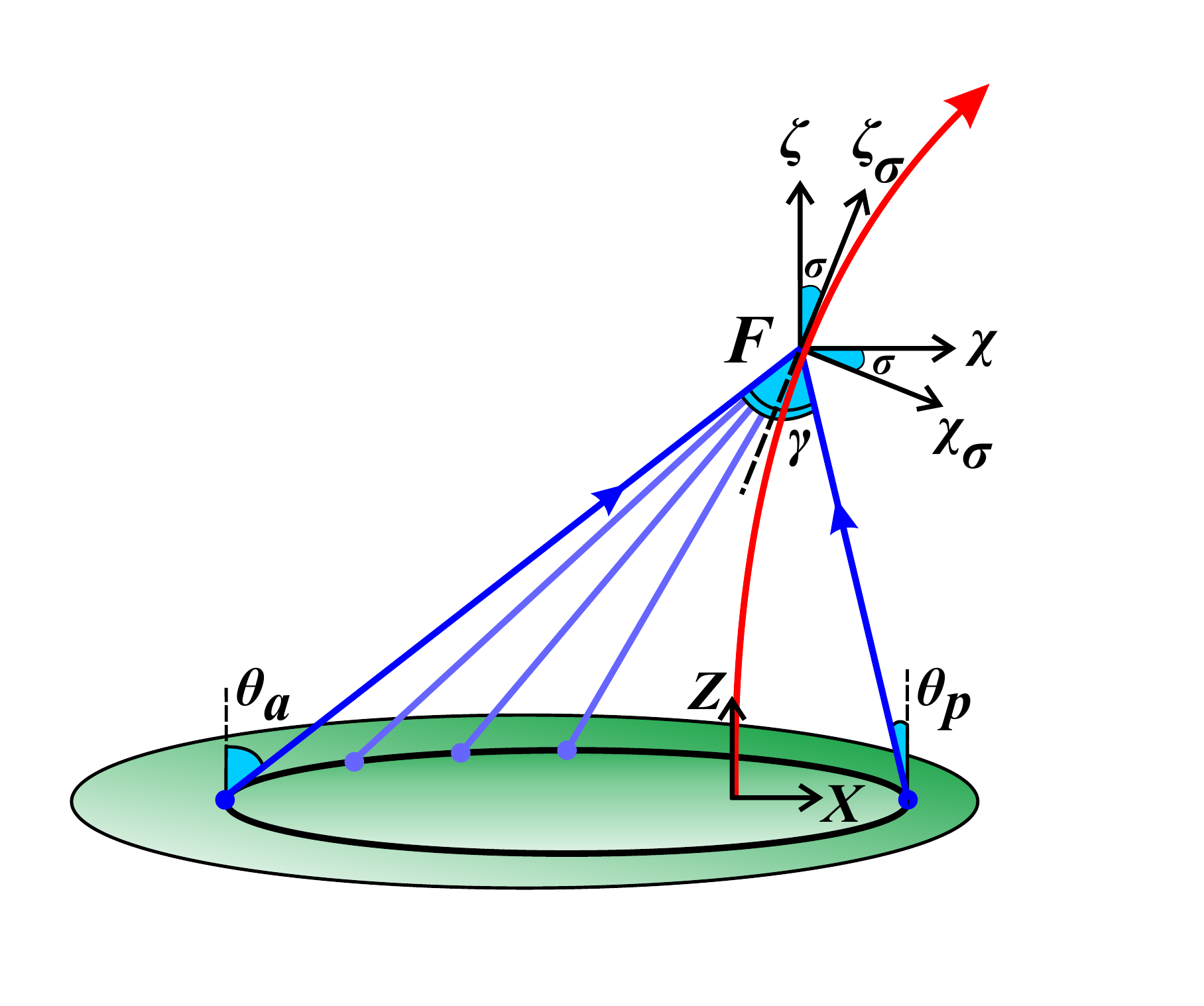}
\caption{Ray optics schematic for the diffraction of the potential function $u$. Rays emitted from expanding ellipses on the input plane interfere to create a curved focal line (red). Shown on the $Y=0$ plane are the global $X, Z$ coordinates, the local $\chi, \zeta$ coordinates at the focal point ${\bf F}(Z)$, the corresponding rotated frame $\chi_{\sigma}, \zeta_{\sigma}$ and the rays from the perigee and apogee points.}
\label{Fig:Figure1}
\end{figure}

Now assume that the rays create a continuous focal line lying on the $Y=0$ plane with a given parametric expression $X=f(Z)$ (see Fig. \ref{Fig:Figure1}). This implies that, at any point ${\bf F}(Z) = (f(Z),0,Z)$ of this line, rays emanating from a geometric locus $C_Z$ on the input plane intersect. The shape of the locus is yet to be determined. Rearranging Eqs. \eqref{Eq:Stationarity} and substituting the equation of the focal line we get
\begin{equation}
\left( {{Q_x},{Q_y}} \right) = \frac{k}{{\left| {{\bf{F}}(Z) - {\bf{r}}} \right|}}\left( {f\left( Z \right) - x, - y} \right)
\label{Eq:PhaseGradient}  
\end{equation}
If every point $\bf r$  on the input plane is mapped to the $Z$ coordinate of the corresponding focal point ${\bf F}(Z)$, a two-variable function $Z({\bf r})=Z(x,y)$  is obtained and the locus $C_Z$  can be interpreted as an \textit{isoline} of this function with value  $Z$. In this way, Eqs. \eqref{Eq:PhaseGradient} give the gradient of the phase $\nabla_{x,y}Q$ at any point $\bf r$.

Proceeding as in \cite{Chremmos_2013_Bessel} one notes that, if Q is to be twice continuously differentiable, its mixed partials should be equal, i.e. $Q_{xy}=Q_{yx}$. After some algebra we obtain from Eqs. \eqref{Eq:PhaseGradient}
\begin{equation}
{Z_x}y\left( {x - f - \frac{Z}{{f'}}} \right) =  - {Z_y}\left[ {{y^2} + \frac{Z}{{f'}}\left( {x - f} \right) + {Z^2}} \right]
\label{Eq:Qxy=Qyx}  
\end{equation}
where the prime denotes the derivative $d/dZ$. The latter equation provides the direction of the gradient $\nabla_{x,y}Z$  which is of course normal to the isoline or ${Z_x}dx + {Z_y}dy = 0$. Therefore it is easy to see that, for any fixed $Z$, $C_Z$ satisfies the differential equation
\begin{equation}
\left[ {{y^2} + \frac{Z}{{f'}}\left( {x - f} \right) + {Z^2}} \right]dx - \left( {x - f - \frac{Z}{{f'}}} \right)ydy = 0
\label{Eq:Differential}  
\end{equation}
The above is solved by multiplying with the integrating factor ${\left( {x - f - {\rm Z}/f'} \right)^{ - 3}}$  and after some manipulation the result reads
\begin{equation}
C_Z : \frac{{{{\left( {x - {x_0}} \right)}^2}}}{{{a^2}}} + \frac{{{y^2}}}{{{b^2}}} = 1
\label{Eq:Ellipse}  
\end{equation}
where
\begin{equation}
a(Z) = Z{\left[ {\left( {w - 1} \right)\left( {1 + w{{f'}^2}} \right)} \right]^{1/2}},\quad b(Z) = Z{\left( {w - 1} \right)^{1/2}},\quad {x_0(Z)} = f - wZf'
\label{Eq:EllipseParameters}  
\end{equation}
and $w(Z)$ is an arbitrary dimensionless function. From Eqs. \eqref{Eq:Ellipse} and \eqref{Eq:EllipseParameters} it follows that $C_Z$ is a conic section, in particular an ellipse for $w>1$ and a hyperbola for $w < -{ ( {f'} )^{-2}}$. Of interest to this work is the first case which generalizes our findings for the paraxial regime \cite{Chremmos_2013_Bessel}: The continuous focal line is created by the apexes of expanding ray cones (\textit{right circular} cones, in particular) whose intersection with the input plane are the ellipses described by Eq. \eqref{Eq:Ellipse}. As shown in Fig. \ref{Fig:Figure1}, the half-angle $\gamma (Z)$ of the cone, for a given $Z$, is easily determined on the $Y=0$ plane as $\gamma = \frac{1}{2}(\theta_a - \theta_p)$, where $\theta_{a,p}$ are, respectively, the angles between the rays from the \textit{apogee} $(x_0-a, 0, 0)$ and \textit{perigee} $(x_0+a, 0, 0)$ points of the ellipse and the $Z$ axis. These are given by $\theta_{a,p}(Z) = {\arctan} \left( {\frac{{f - {x_0} \pm a}}{Z}} \right) $ and after some trigonometry the result reads   
\begin{equation}
\gamma  = \arctan{\left( {\frac{{w - 1}}{{1 + w{{f'}^2}}}} \right)^{1/2}}
\label{Eq:HalfAngle}  
\end{equation}
Obviously $\tan \gamma < | f'|^{-1}$ or equivalently $\gamma + |\sigma| < \pi /2$ with $\sigma = \arctan f' $ which physically implies that the most inclined ray (the one from the apogee point) cannot be vertical to the $Z$ axis. A nonparaxial beam implies that the sum $\gamma + |\sigma|$ assumes large values. This in turn implies that either the cone half-angle (which is related to the width of a Bessel beam) or the angle of the trajectory or both assume large values. In the other extreme, namely the limit of focal trajectories with small slopes $( |f'| \ll 1 )$  and small cone angles $(\gamma \ll 1)$, one has $w \approx 1 + {\gamma ^2}$  and Eq. \eqref{Eq:Ellipse} reduces to ${\left( {x - f + Zf'} \right)^2} + {y^2} = {\gamma ^2}{Z^2}$  which is exactly the circle equation that was obtained in the paraxial case \cite{Chremmos_2013_Bessel}.

Another interesting result concerns the angle between the axis of the ray cone from $C_Z$ and the $Z$ axis. This can again be found on the $Y=0$ plane from the bisector of the cone angle and reads $\gamma + \theta_p = \frac{1}{2}(\theta_a + \theta_p)$ which, after some similar trigonometric steps, yields $\sigma$. Therefore the axis of the ray cone is \textit{tangent} to the focal curve at the current focal point ${\bf F}(Z)$. This further suggests that, if the optical field is observed on the plane that passes from ${\bf F}(Z)$ and is normal to the tangent to the focal trajectory (plane $\zeta_{\sigma} = 0$ in Fig. \ref{Fig:Figure1}), a conical superposition of rays occurs with angle $2\gamma$, thus leading (approximately) to a $J_0(k_{\bot} {\rho _\sigma})$ distribution of the field around the focus, with $\rho _\sigma$ being the local polar-distance variable and $k_{\bot}=k \sin \gamma$ the projection of the wave vector on that plane. Obviously, $\gamma$ has to stay constant with $Z$ if the beam is to sustain a quasi-diffraction-free profile on this plane.

Summarizing our findings, the design of a continuous focal line in the general non-paraxial case has led us to beams with a Bessel-like profile on the plane that stays constantly normal to the trajectory of the focus. This is a central result of this paper that generalizes the recently reported case of paraxial accelerating Bessel-like beams whose profile is invariant on $Z=ct.$ planes \cite{Chremmos_2013_Bessel}. It should also be noted that this result further suggests the possibility to extend or modify the definition of accelerating diffraction-free beams in the paraxial regime to include generic waves that stay propagation invariant in a coordinate frame of \textit{self-reference}, i.e. on a plane that stays normal to the beam's trajectory. While keeping this in mind, we will here restrict ourselves to Bessel-like beams, which we consider of high practical interest, and leave the theoretical investigation of this possibility for a follow-up work.

\subsection{Focal field distribution}

Let us now compute the distribution of the optical field in the neighborhood of the focal point ${\bf F}(Z)$ at the fixed distance $Z$. To this end we introduce on the input plane new dimensionless elliptic coordinates $(\mu,\nu)$ around the center $(x_0,0)$ of the ellipse $C_Z$: $x = x_0 + a\mu \cos \nu$, $y = b\mu \sin \nu$; a constant $\mu >0$ corresponds to ellipses with constant eccentricity while a constant $\nu \in [0, 2\pi)$ corresponds to straight lines passing from $(x_0,0)$. Then we have from Eq. \eqref{Eq:Rayleigh-Sommerfeld}
\begin{equation}
u\left( {\bf{F}} \right) =  - \frac{{k^3 abZ}}{{2\pi }}\int\limits_0^{2\pi } {\left( {\int\limits_0^\infty  {U\left( {\mu ,\nu } \right)g\left( k{\left| {{\bf{F}} - {\bf{r}}\left( {\mu ,\nu } \right)} \right|} \right){e^{iP\left( {{\bf{F}},\mu ,\nu } \right)}}\mu d\mu } } \right)d\nu }
\label{Eq:Rayleigh-Sommerfeld-alter}  
\end{equation}
where $U(\mu, \nu)=U({\bf r}(\mu, \nu))$ and $P({\bf F}, \mu, \nu)=P({\bf F}, {\bf r}(\mu, \nu))$. Now note that $\mu =1$ corresponds to the ellipse $C_Z$, hence $P_{\mu}({\bf F}, 1, \nu)=0$ for all $\nu$, which suggests a stationary-phase approximation to the inner integral. We obtain  
\begin{equation}
u\left( {{\bf{F}} + \delta {\bf{R}}} \right) \approx  - \frac{{k^3 abZ{e^{i\pi /4}}}}{{{{\left( {2\pi } \right)}^{1/2}}}}\int\limits_0^{2\pi } {\frac{{U\left( {1,\nu } \right)g\left( k{\left| {{\bf{F}} - {\bf{r}}\left( {1,\nu } \right)} \right|} \right)}}{{P_{\mu \mu }^{1/2}\left( {{\bf{F}},1,\nu } \right)}}{{\rm{e}}^{iP\left( {{\bf{F}},1,\nu } \right) + i{\nabla _{\bf{R}}}P\left( {{\bf{F}},1,\nu } \right)\cdot\delta {\bf{R}}}}d\nu }
\label{Eq:Rayleigh-Sommerfeld-Approx_1}  
\end{equation}
where we have also accounted for a small displacement $\delta {\bf R}=(\chi, \psi, \zeta)$ around the focus by Taylor-expanding the phase $P$ at the point $({\bf F}, 1, \nu)$. Only the $\nabla _{\bf R}$ derivatives appear since $P_{\mu}=P_{\nu}=0$ at this point. Moreover, stationarity implies that the value of the phase is constant along $C_{Z}$, i.e. $P(Z) \triangleq P({\bf F}, 1, \nu)$, and thus $\exp(iP)$ can be pulled out of the integral. Function $P(Z)$ will be determined in the following. Note also that, for simplicity, the variations of the wave amplitude have been neglected for displacements of few $\lambda$ around the focus. The extra phase due to the displacement of the observation point from the focus is written explicitly 
\begin{equation}
{\nabla _{\bf{R}}}P\left( {{\bf{F}},1,\nu } \right) \cdot \delta {\bf{R}} = k\frac{{{\bf{F}} - {\bf{r}}\left( {1,\nu } \right)}}{{\left| {{\bf{F}} - {\bf{r}}\left( {1,\nu } \right)} \right|}} \cdot \delta {\bf{R}} = {\bf{k}}\left( {1,\nu } \right) \cdot \delta {\bf{R}}
\label{Eq:ExtraPhase}  
\end{equation}
which implies that the ray from point ${\bf r}(1,\nu)$ of $C_Z$ contributes (approximately) in the neighbourhood of ${\bf F}(Z)$ a plane wave with wave vector ${\bf k}(1,\nu)$. Next, to find $P_{\mu \mu}$ we apply the chain rule to the composite function $P({\bf R},{\bf r}(\mu, \nu))$. Since $P_x=P_y=0$, only the second-order partial derivatives $P_{xx}$, $P_{xy}$ and $P_{yy}$ are involved, which are straightforward to compute from the definition of $P({\bf F}, {\bf r})$ (where $\bf F$ is fixed). Also required are the corresponding second-order derivatives of $Q$ which are obtained from Eq. \eqref{Eq:PhaseGradient}, keeping in mind that $Z$ is a function of $x$, $y$. The needed gradient $\nabla_{x,y}Z$ can be obtained from Eq. \eqref{Eq:Ellipse} by differentiating with respect to $x$ and $y$ and reads
\begin{eqnarray}
\left( {{Z_x},{Z_y}} \right) &=& \left( {\frac{b}{{Da}}\left( {x - {x_0}} \right),\frac{a}{{Db}}y} \right), \nonumber \\
\text{where }D &=& \frac{{a'b}}{{{a^2}}}{\left( {x - {x_0}} \right)^2} + \frac{{ab'}}{{{b^2}}}{y^2} + \frac{{b{x'_0}}}{a}\left( {x - {x_0}} \right)
\label{Eq:ZGradient}
\end{eqnarray}
Combining all the results and using also Eqs. \eqref{Eq:EllipseParameters}, we obtain after some long algebra the remarkably simple result
\begin{eqnarray}
{P_{\mu \mu }}({\bf{F}},1,\nu ) &=& {P_{xx}}x_\mu ^2 + 2{P_{xy}}{x_\mu }{y_\mu } + {P_{yy}}y_\mu ^2 \nonumber \\
 &=& \frac{{k{a^2}{b^2}}}{{{w^{1/2}}D(1,\nu ){{\left| {{\bf{F}} - {\bf{r}}(1,\nu )} \right|}^2}}}
\label{Eq:Chain_Rule}
\end{eqnarray}
where $D = a'b{\cos ^2}\nu  + ab'{\sin ^2}\nu  + b{x'_0}\cos \nu$ follows from the definition of Eq. \eqref{Eq:ZGradient}.

Now in order to see how Eq. \eqref{Eq:Rayleigh-Sommerfeld-Approx_1} expresses a conical superposition of waves, we replace the ``elliptic azimuth" variable $\nu$ (which determines the position of $\bf r (1,\nu)$ along the ellipse $C_Z$) by the actual azimuth angle $\phi$ with respect to the axis of the corresponding cone. This is accomplished through the transformation
\begin{equation}
\cos \nu = \frac{f' \tan \gamma + \cos \phi}{1 + f' \tan \gamma \cos \phi}
\label{Eq:Transformation}
\end{equation}
where $\phi \in [0, 2\pi)$. It also follows that the differentials $d\nu$, $d\phi$ are related through $aZd\nu  = |{\bf{F}} - {\bf{r}}(1,\nu )|b\cos \gamma d\phi$. Introducing the new integration variable $\phi$ into Eq. \eqref{Eq:Rayleigh-Sommerfeld-Approx_1} and with the help of Eqs. \eqref{Eq:ExtraPhase} and \eqref{Eq:Chain_Rule} we obtain
\begin{align}
u\left( {{\bf{F}} + \delta {\bf R}_{\sigma}} \right) & \approx  - {\left( {\frac{{ik}}{{2\pi }}} \right)^{1/2}}C\left( Z \right){{\rm{e}}^{i\left[ {P\left( Z \right) + {k_{\parallel}}{\zeta _\sigma }} \right]}} \nonumber \\
 & \times \int\limits_0^{2\pi } {U\left( {1,\phi } \right){D^{1/2}}\left( {1,\phi } \right)\tilde g\left( {k\left| {{\bf{F}} - {\bf{r}}\left( {1,\phi } \right)} \right|} \right){{\rm{e}}^{ - i{k_ \bot }{\rho _\sigma }\cos \left( {\phi  - \theta } \right)}}d\phi }
\label{Eq:Rayleigh-Sommerfeld-Approx_2}
\end{align}
where $C\left( Z \right) = \left( \cos(\gamma + \sigma) \cos(\gamma - \sigma) \right)^{1/4} (\cos \sigma)^{1/2}$, $\tilde g (r)= {r^2}g(r)$, $k_{\parallel}=k \cos \gamma$, $k_{\bot}=k \sin \gamma$, and 
\begin{equation} 
\delta {\bf R}_{\sigma} = \left( \chi_{\sigma},\psi_{\sigma},\zeta_{\sigma} \right) = \left( {\chi \cos \sigma - \zeta \sin \sigma,\psi,\chi \sin \sigma + \zeta \cos \sigma} \right)
\label{Eq:RotatedFrame}
\end{equation}
The above is a new frame of displacement coordinates that is obtained by rotating the frame $(\chi, \psi, \zeta)$ around the $\psi$ axis by $\sigma$ radians so that the $\zeta _{\sigma}$ axis coincides with the axis of the current ray cone (see \ref{Fig:Figure1}), while $(\rho_{\sigma}, \theta_{\sigma}, \zeta_{\sigma})$ are the corresponding cylindrical coordinates with $\rho_{\sigma} = (\chi_{\sigma}^2 + \psi_{\sigma}^2)^{1/2}$ and $\theta_{\sigma} = \arctan(\psi_{\sigma} / \chi_{\sigma})$. Equation \eqref{Eq:Rayleigh-Sommerfeld-Approx_2} clearly shows that the field around the focus results from a superposition of a conical bundle of plane waves at an angle $\gamma$ with the cone axis. These waves have a common wave vector component parallel to the cone axis equal to $k_{\parallel}$, resulting in the common phase $k_{\parallel} \zeta _{\sigma}$, while their projections on the $\zeta _{\sigma} = 0$ plane form an angular spectrum of plane waves with wave vector magnitude $k_{\bot}$ and an amplitude that varies azimuthially with $\phi$ as $UD^{1/2}\tilde g$.

To complete the computation of the focal field one has to determine the \textit{focal phase} function $P(Z)$. This is obtained directly from the definition $P(Z) = Q({\bf{r}}) + k \left| {{\bf F}(Z) - {\bf r}} \right|$ by differentiating with respect to $x$ or $y$ (either will do) and using Eq. \eqref{Eq:PhaseGradient}. The result is
\begin{equation} 
P\left( Z \right) = k\int\limits_0^Z {\frac{{\cos \left( {\gamma \left( z \right)} \right)}}{{\cos \left( {\sigma \left( z \right)} \right)}}dz}
\label{Eq:P}
\end{equation}
Noting that $dz/\cos \sigma  = dz{\left( 1 + {f'}^2 \right)^{1/2}}$ is the infinitesimal length along the focal curve $X=f(Z)$, Eq. \eqref{Eq:P} has a very clear interpretation: the phase of the focal field at ${\bf F}(Z)$ is equal to the line integral of the tangential wave vector $k_{\parallel}$ along the focal trajectory. 

Let us now see how Eq. \eqref{Eq:Rayleigh-Sommerfeld-Approx_2} can yield a Bessel-like profile. In propagation distances of a few wavelengths $k|{\bf F} - {\bf r}| \gg 1$, thus one can approximate $\tilde g \approx i$. Furthermore, some additional manipulation of $D$ from Eq. \eqref{Eq:ZGradient} with the help of Eqs. \eqref{Eq:EllipseParameters} yields
\begin{equation} 
D(1, \phi) = \frac{{{\rho _\sigma }\left( {1,\phi } \right)}}{{2w{C^2}\left( Z \right)}}\frac{{{{\cos }^2}\sigma }}{{\cos \gamma }}\left\{ {\frac{{{{\left[ {{Z^2}\left( {w - 1} \right)} \right]}^\prime }}}{{Z\tan \gamma }} - \frac{{Z{{\left( {w{{\tan }^2}\sigma } \right)}^\prime }}}{{\tan \sigma }}\cos \nu} \right\}
\label{Eq:D}
\end{equation}
where $\rho_{\sigma}(1,\phi)$ is the polar distance of point ${\bf r }(1,\phi)$ with respect to the axis of the current ray cone and variable $\nu$ has been deliberately been used (instead of $\phi$) in the right-hand side. Now assume that the focal trajectory is straight $(\sigma ' = 0)$ and that the angle of the ray cone is constant $(\gamma ' = 0)$. From Eq. \eqref{Eq:HalfAngle}, it also follows that $w' = 0$. In this case Eqs. \eqref{Eq:P} and Eq. \eqref{Eq:D} give respectively $P = kZ\cos \gamma / \cos \sigma$ and $D = \rho _{\sigma} \sin \gamma / C^2$. If we additionally assume that the input amplitude satisfies $U\left( {1,\varphi } \right) = {\left( {2\pi {k_ \bot }{\rho _\sigma }} \right)^{ - 1/2}}{{\rm{e}}^{i\pi /4}}$, Eq. \eqref{Eq:Rayleigh-Sommerfeld-Approx_2} yields 
\begin{equation} 
u\left( {{\bf{F}} + \delta {{\bf{R}}_\sigma }} \right) \approx {J_0}\left( {{k_ \bot }{\rho _\sigma }} \right)\exp \left[ {i{k_{\parallel}}\left( {\frac{Z}{{\cos \sigma }} + {\zeta _\sigma }} \right)} \right]
\label{Eq:Bessel_Field}
\end{equation}
where the familiar representation of the $J_0$ Bessel function as a uniform angular spectrum of plane waves has been used. Equation \eqref{Eq:Bessel_Field} clearly represents an ideal diffraction-free Bessel beam that propagates with wave vector $k_{\parallel}$ along the axis $X = Z \tan {\sigma}$. Also, the selection of the input amplitude $U$ as proportional to $\rho _{\sigma} ^{-1/2}$ (where $\rho _{\sigma}$ is the distance from the axis of the beam), agrees with the well known asymptotic expression of the Bessel function for large arguments. Therefore the ray optics approximation provided by Eq. \eqref{Eq:Rayleigh-Sommerfeld-Approx_2} is consistent with the fact that the Helmholtz equation accepts Bessel beam solutions.

What is however far more interesting about Eq. \eqref{Eq:Rayleigh-Sommerfeld-Approx_2} is that, even when the trajectory is curved $(\sigma ' \neq 0)$, the superposition of the angular plane-wave spectrum occurs always on the $\zeta _{\sigma} = 0$ plane, namely the normal to the trajectory of the focus plane. It is thus possible to sustain a quasi-diffraction-free Bessel profile $J_0(k_{\bot} \rho _{\sigma})$ on that dynamically changing plane provided that the cone angle $\gamma$ is kept constant with $Z$ and that the input wave amplitude varies as $U \propto C^{-1}D^{-1/2}$. The first condition is easily imposed as $\gamma$ enters directly the algorithm for computing the input phase $Q$ as will be described in the following. The condition for the amplitude is very difficult to implement in practice and one is usually restricted to simple optical envelopes such as plane wave or Gaussians. Nevertheless, as our numerical experiments show, using a Gaussian envelope does not impair significantly the main features of the beam, as long as the input phase is correctly engineered. Namely, the beam profile will approximately follow Eq. \eqref{Eq:Bessel_Field} with the linear phase replaced by the phase accumulated along the curved trajectory, i.e. $ {P\left( Z \right) + {k_{\parallel}}{\zeta _\sigma }}$.

Another interesting remark concerns the profile of the beam on the $\zeta = 0$ plane, which is the standard, parallel to the input, plane on which optical beams are usually observed. Using Eqs. \eqref{Eq:RotatedFrame} we get ${J_0}\left( {{k_ \bot }\sqrt {{{\left( {\chi \cos \sigma } \right)}^2} + {\psi ^2}} } \right)$, which is an elliptically deformed Bessel profile. The eccentricity of the elliptic isolines of this profile is $\sin \sigma$, i.e. it increases with the local slope of the trajectory. 

\subsection{Final design algorithm}
\label{Section: Algorithm}

The procedure to determine the input phase of the beam is now clear: The phase at any point $\bf r$ on the input plane is given by $Q({\bf r}) = P(Z) - k|{\bf F}(Z)-{\bf r}|$ where $Z$ corresponds to the locus $C_Z$ passing from this point. To find this locus, one has to solve the nonlinear equation \eqref{Eq:Ellipse} for $Z$ setting $x,y$ equal to the coordinates of $\bf r$. A critical point in this procedure is that the solution for $Z$ must be \textit{unique}, namely there must be no intersection of the loci $C_Z$ for different $Z$. As explained in \cite{Chremmos_2013_Bessel}, this is equivalent to the requirement that $\nabla Z$ stays finite. From Eq. \eqref{Eq:ZGradient}, this implies $D \neq 0$ or more specifically $D > 0$, which implies an outward gradient (expanding ellipses). Particularly helpful at this point is the alternative expression \eqref{Eq:D}, which shows immediately that the first term inside the brackets must be greater than the absolute value of the factor of $\cos \nu$ (thus justifying our choice to use $\nu$ instead of $\phi$ in this expression). After some algebraic manipulations we end up with the remarkably simple inequality
\begin{equation} 
| Z\sigma' - \sin^2\gamma \tan\sigma | < Z\gamma' + \frac{1}{2}\sin2\gamma
\label{Eq:Inequality}
\end{equation}
A first interesting conclusion is that, since $|\sigma|+\gamma < \pi/2$, the above condition is true for all $Z$ if $\gamma' > |\sigma'|$. Therefore the ellipses never intersect if the cone half-angle increases with $Z$ faster than the increase or decrease rate of the angle of the trajectory with the $Z$ axis. More interesting is though the case of constant $\gamma$ corresponding to a diffraction-resisting beam. Then, for any given function $\sigma (Z) = \arctan{f'(Z)}$, the inequality \eqref{Eq:Inequality} defines a maximum propagation distance $Z_m$ at which the focal curve $X=f(Z)$ can be created or, equivalently, a maximum ellipse $C_{Z_m}$ in the exterior of which the computation of $Q$ fails. In addition to \eqref{Eq:Inequality}, $Z_m$ must satisfy $|\sigma| + \gamma < \pi/2$ as explained below \eqref{Eq:HalfAngle}; hence $Z_m$ is actually the minimum $Z$ satisfying both conditions. Beyond $Z_m$, a different trajectory must be defined, with the simplest choice being $Z = f(Z_m) + f'(Z_m)(Z-Z_m)$, namely the line tangent to the trajectory at the ultimate point ${\bf F}(Z_m)$ \cite{Chremmos_2013_Bessel}.

It is also interesting to see how \eqref{Eq:Inequality} simplifies in the paraxial limit. In this case $\sigma \approx f' \ll 1$, $\gamma \ll 1$, $a \approx b \approx \gamma Z$ (the ellipses become circles) and \eqref{Eq:Inequality} reduces to $Zf'' < (Z\gamma)' = b'$ which is exactly the condition obtained in \cite{Chremmos_2013_Bessel}.

\section{Numerical experiments}

In this Section the proposed beams are demonstrated through a series of numerical simulations. In all cases, the envelope of the input condition at $Z=0$ is assumed in the simple form of a Gaussian with $e^{-1}$ width $\alpha$, namely $U(x,y)=\exp(-(x^2+y^2)/\alpha^2)$. Given the trajectory function $X=f(Z)$ and the width of the beam's main lobe (determined by the cone half angle $\gamma$), the input phase is obtained from the procedure of paragraph \ref{Section: Algorithm}. Subsequently the Gaussian envelope is multiplied with the phase mask $\exp(iQ(x,y))$ and the resulting wavefront $u(x,y,0)$ is allowed to propagate in $Z>0$ according to the Helmholtz equation. Finally, the EM field components are computed from \eqref{Eq:Potential}.

\begin{figure}[]
\centering
\includegraphics[width=1.0\textwidth]{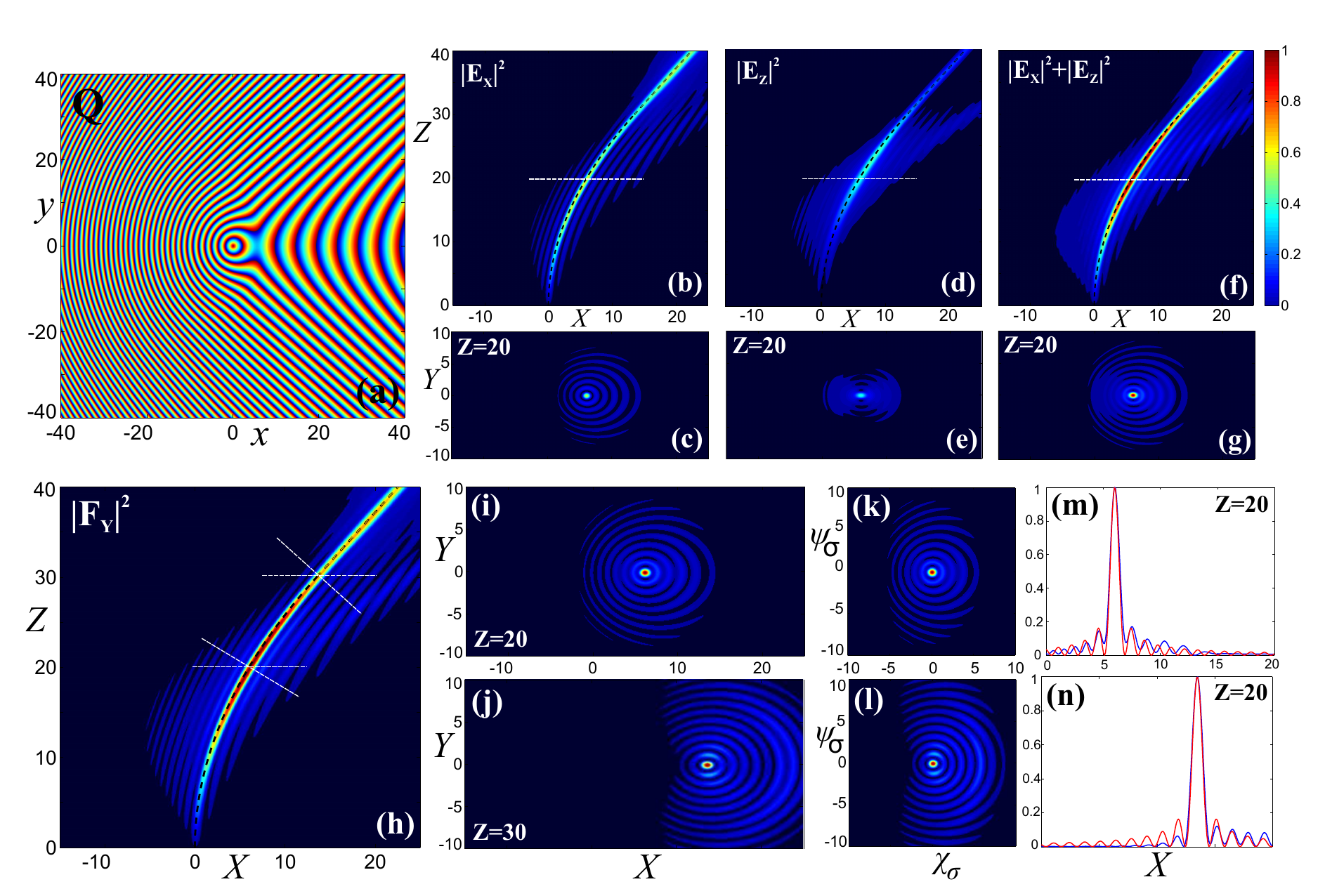}
\caption{Simulation results for a beam with $f(Z)=0.015Z^2$, $\gamma = 30^0$, $Z_m = 30$ and $\alpha = 50$. (a) Input phase Q modulo $2 \pi$. (b) Evolution of $|E_X|^2$ on the plane $Y=0$ and (c) its transverse profile on the plane $Z=20$ indicated in (b) with white dashed line. (d-e) The same for $|E_Z|^2$. (f-g) The same for the electric energy density $|E_X|^2 + |E_Z|^2$. The black dashed curves indicate the theoretical trajectory. The maps have the same colour code to allow comparison of the intensities. (h) Evolution of $|F_Y|^2$ on $Y=0$, (i,j) its transverse profiles at $Z=20$ and $Z=30$ and (k-l) profiles at the corresponding normal to the trajectory $\chi_\sigma - \psi_\sigma$ planes indicated in (h) with white dashed line. (m,n) Comparison of the simulated (blue) and theoretical (red) curves for $|F_Y|^2$ on $Z=20$ and $Z=30$. All distances are measured in wavelengths.}
\label{Fig:Figure2}
\end{figure}

Figure \ref{Fig:Figure2} demonstrates a paraxial Bessel-like beam with a parabolic trajectory of the form $f(Z)=\kappa Z^2$. The simulation parameters were taken $\kappa = 0.015$, $\gamma = 30^0$, $Z_m = 30$ and $\alpha = 50$. All distances are measured in wavelengths. For a standard non-accelerating Bessel beam, this $\gamma$ value corresponds to a subwavelength FWHM of the central lobe equal to $2.25/k_\bot \approx 0.72\lambda$. For $Z>Z_m$ the trajectory is continued along its tangent at the ultimate point $Z_m$. The maximum angle of this trajectory with respect to the $Z$ axis is $\sigma_m = \arctan(2 \kappa Z_m) \approx 42^0$. This in turn results in a maximum inclination of the involved rays, with respect to the $Z$ axis, equal to $\gamma + \sigma_m \approx 72^0$, indicating the clear nonparaxial character of the wave. From the figure, the agreement of the trajectory of the simulated beam with the theoretical is obvious. Note also in (c) the deformed Bessel-like profile of the stronger $E_X$ component which follows from the corresponding profile of the vector potential shown in (i). The longitudinal $E_Z$ component in (e) is weaker but its magnitude increases to become comparable to $E_X$ as the trajectory gets more inclined. This is totally reasonable since the wave loses its transverse nature the more it departs from the paraxial regime. By closer examination, one sees that $E_Z$ initially has a node (due to the derivative $\partial u/\partial X$) which progressively disappears due to the linear phase factor that multiplies the Bessel function in \eqref{Eq:Bessel_Field}. Overall, the presence of the longitudinal component does not affect the Bessel profile significantly, as seen from the total electric energy density shown in (g). Hence the beam's energy profile inherits the Bessel-like form of the electric vector potential. To further verify our predictions, subfigures (h-n) depict the vector potential. In (i) and (j) the profiles along $Z=20$ and $Z=30$ are shown respectively. The elongation of the profile along $X$ increases as predicted by theory. In (m) and (n) the intensity on these planes and along the line $Y=0$ is plotted in comparison to the theoretical curve $J_0^2\left( {{k_ \bot }\left| {X - f\left( Z \right)} \right|\cos \sigma } \right)$ and the agreement is very satisfactory. The FWHM of these curves is approximately 0.9 for (m) and 1 for (n), which is very close to the theoretical values 0.84 and 0.96 obtained from the formula $2.25/(k_\bot \cos \sigma)$. Also, in (k) and (l), the profiles of $|F_Y|^2$ along the normal to the trajectory planes at $Z=20$ and $Z=30$ are shown and verify the expected circularly symmetric ring structure. The electric energy density behaves similarly. Hence our prediction is verified that the beam exhibits a diffraction-resisting property on a dynamically changing plane that stays normal to the trajectory of the beam.  

\begin{figure}[h]
\centering
\includegraphics[width=0.9\textwidth]{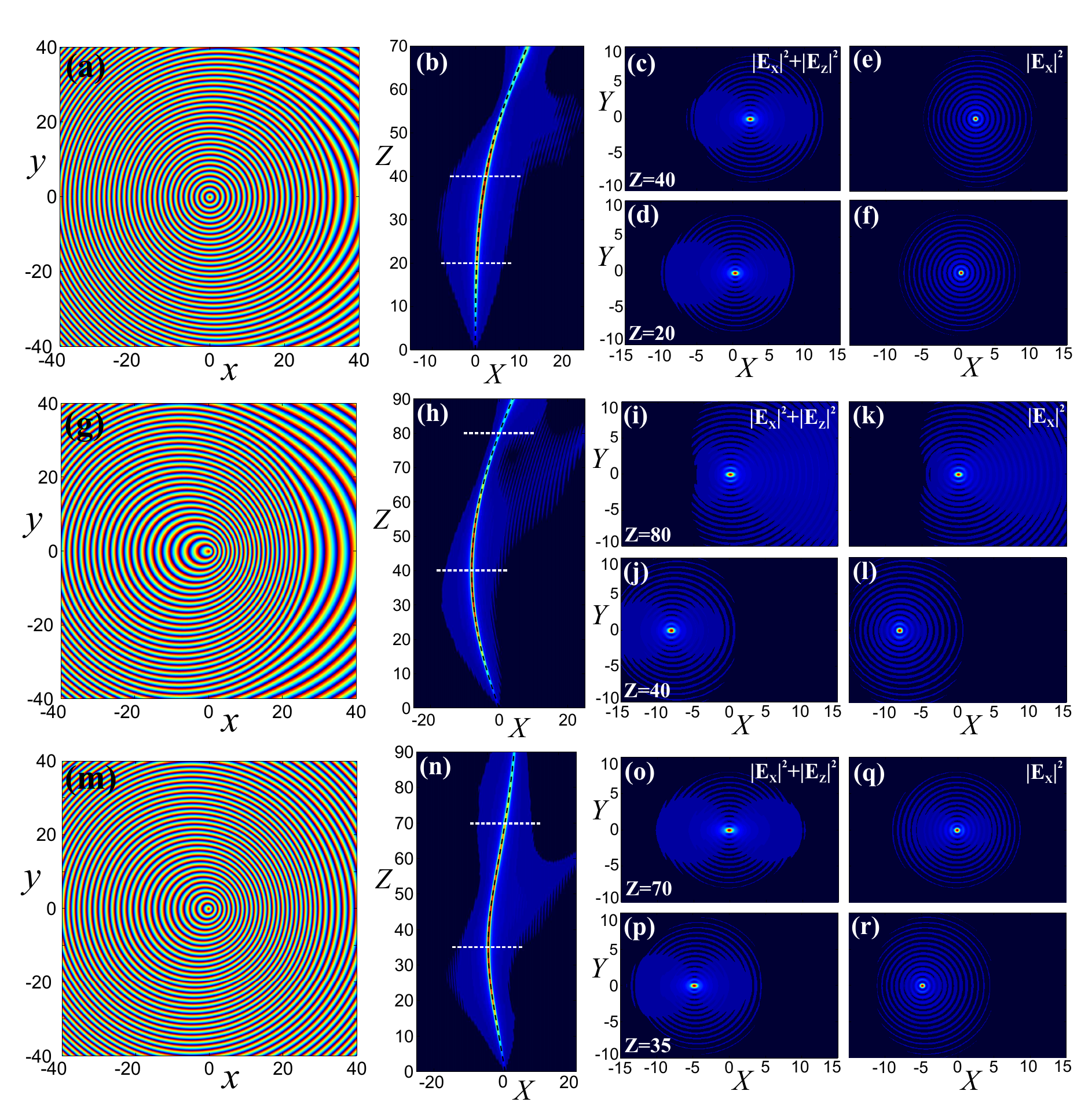}
\caption{Top row: Results for a beam with $f(Z)=(3.5 \times 10^{-5})Z^3$, $\gamma = 45^0$, $Z_m = 60$ and $\alpha = 80$: (a) Input phase Q modulo $2 \pi$. (b) Evolution of electric energy density $|E_X|^2 + |E_Z|^2$ on the plane $Y=0$ and its transverse profile on the planes (c) $Z=20$ and (d) $Z=40$, indicated in (b) with white dashed lines. (e-f) Corresponding profile for the $E_X$ component only. The black dashed curve in (b) indicates the theoretical trajectory. Middle row: Corresponding results for a beam with $f(Z)=0.66\left( {\sqrt {{Z^2} - 80Z + 5200}  - 20\sqrt {13} } \right)$, $\gamma = 40^0$, $Z_m = 90$ and $\alpha = 80$. Bottom row: Corresponding results for a beam with $f(Z) =  - 10 \text{sech}[0.036(Z - 35)]$, $\gamma = 45^0$, $Z_m = 100$ and $\alpha = 80$. All distances are measured in wavelengths.}
\label{Fig:Figure3}
\end{figure}

It may also be interesting to note that, for parabolic trajectories, the inequality \eqref{Eq:Inequality} leads to an interesting analytical condition: For beams with $\gamma > \arctan \sqrt{5/27} \approx 23^0$ the inequality holds for all $Z$, hence the maximum distance $Z_m$ is bounded only by the condition $\sigma(Z) + \gamma < \pi/2$. For smaller angles $\gamma$, \eqref{Eq:Inequality} leads to a cubic equation for $Z$ which can also be solved analytically. Higher power-law or transcendental trajectories can though be treated only numerically.

A set of different trajectories are investigated in \ref{Fig:Figure3}. In the top row, a Bessel-like beam with a cubic trajectory is shown. Due to the quadratically increasing with $Z$ acceleration, the cone half-angle has to be quite large so that the condition \eqref{Eq:Inequality} is satisfied up to a distance of few tens of wavelengths. Here $\gamma = 45^0$ which corresponds to a subwavelength main lobe with FWHM around $\lambda / 2$ which clearly suggests a non-paraxial wave. The maximum distance to which the beam accelerates is $60\lambda$ and the corresponding maximum angle of the trajectory with respect to the $Z$ axis is $\sigma_m=21.8^0$ also adding to the nonparaxiality of the wave. The agreement of the trajectory of the simulated beam with the theoretical curve is excellent as shown in (b). Remarkable is also the nearly perfect ring pattern of the snapshots of $|E_X|^2$ at different transverse planes shown in (e-f). The total energy of (c-d) has a similar pattern which is however distorted and elliptically deformed due to the "dipolar" distribution of the energy of the $E_Z$ component (not shown). Despite this distortion the beam can serve well as a diffraction-resisting nonparaxial accelerating wave. 

The middle and bottom row of Fig. \ref{Fig:Figure3} examine the cases of a hyperbolic and a hyperbolic secant trajectory, respectively. In these examples too, the beams have subwavelength main lobes with FWHM around 0.56 ($\gamma = 40^0$) and 0.51 ($\gamma = 45^0$) wavelengths respectively. Moreover, the maximum bending angles of the trajectories are also nonparaxial: approximately $40.2^0$ for the hyperbolic beam (measured as the difference of the angles at $Z=0$ and $Z=Z_m$) and $20.4^0$ for the hyperbolic secant beam (measured as the difference of the angles at the two inflection points $Z=10.5$ and $Z=59.5$). Again, the agreement of the obtained trajectories with the desired ones is excellent while similar with the cubic case are the comments for the energy density images.

Through the previous examples, the trade-off between the design parameters of the introduced beams is evident. Higher rates of acceleration require larger cone angles (in order to avoid intersection of the ellipses of \eqref{Eq:Ellipse}) which in turn limit the maximum bending angles of the trajectories. This is the price one has to pay for accelerating diffraction-resisting beams based on conical interference.

\begin{figure}[t]
\centering
\includegraphics[width=1.0\textwidth]{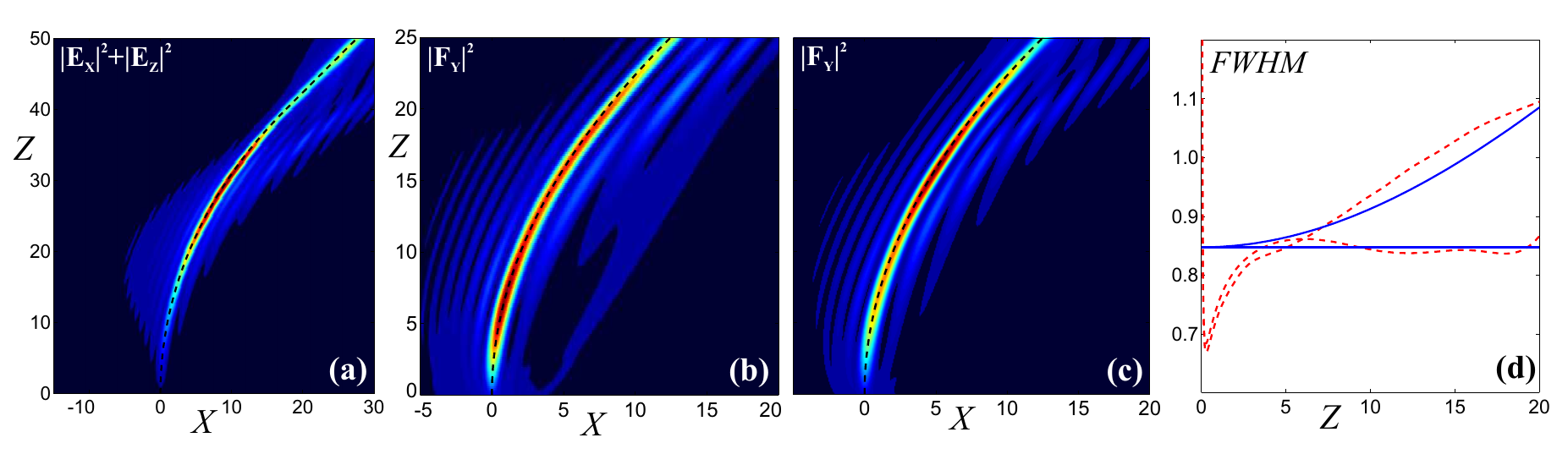}
\caption{(a) Evolution of electric energy density $|E_X|^2 + |E_Z|^2$ on the plane $Y=0$ for a beam with $f\left( Z \right) =  - \left( {160/\pi } \right)\ln \left[ {\cos \left( {\pi Z/160} \right)} \right]$, $\gamma = 20 + 0.25Z (^0)$, $Z_m = 40$ and $\alpha = 60$. (b-c) Evolution on $Y=0$ of the squared modulus of the vector potential of a beam with $f(Z)=0.03Z^2$ and (b) fixed $\gamma = 25^0$ and (c) $\gamma \left( Z \right) = {\sin ^{ - 1}}\left[ {\sin \left( {\gamma \left( 0 \right)} \right)/\cos \left( {\sigma \left( Z \right)} \right)} \right]$. The black dashed curves indicate the theoretical trajectories. (d) Comparison of the FWHM of the beams of (b) and (c) (red dashed lines) with the theoretical curves (blue solid lines) superposed. All distances are measured in wavelengths.}
\label{Fig:Figure4}
\end{figure}

So far, the angle $2\gamma$ of the expanding ray cones remained constant with $Z$. As a result, the beams maintained a diffraction-resisting property when observed on a hypothetical screen that stays normal to the trajectory at any point. In applications where this requirement can be relaxed, e.g. when we are interested more in guiding light along a given trajectory than in the above property of a $Z$-independent cone angle, the possibility of a varying $\gamma$ provides the designer with an additional degree of freedom. An interesting case is to choose the function $\gamma(Z)$ so as to extend the maximum distance $Z_m$ of validity of the condition \eqref{Eq:Inequality}. For example consider the trajectory $f\left( Z \right) =  - \left( {160/\pi } \right)\ln \left[ {\cos \left( {\pi Z/160} \right)} \right]$, whose angle with the $Z$ axis varies linearly with $Z$, as $\sigma(Z)=9Z/8$ $(^0)$. Substituting this function into \eqref{Eq:Inequality} with $\gamma = 20^0$, we find that the inequality holds only for $Z<18.85$, beyond which the ray cones begin to intersect. If however $\gamma$ is allowed to increase with $Z$, the right-hand side of the inequality increases due to the term $Z\gamma'$ and $Z_m$ can be extended significantly. Indeed, by letting $\gamma = 20 + 0.25Z$ $(^0)$, we find that \eqref{Eq:Inequality} holds up to around $Z = 50$, a bound set by the second condition $\gamma + |\sigma| < \pi/2$. The simulated propagation of this beam for $Z_m=40$ is shown in Fig. \ref{Fig:Figure4}(a) in terms of the electric energy density. The agreement of the trajectory with the theoretical one is excellent.

Another useful application of a varying cone angle is the possibility to design a beam whose main lobe stays invariant on planes of constant $Z$. This is understood if one considers the result of the ray optics analysis for the FWHM of the main lobe in the direction of acceleration, namely $2.25{(k\sin \gamma \cos \sigma )^{-1}}$. The latter formula shows that $\gamma(Z)$ can be used to cancel the elongation of the profile which is propotional to the factor $(\cos\sigma)^{-1}$ or equivalently $\sqrt {1 + {{\left( {f'} \right)}^2}}$. This is obviously achieved by choosing $\gamma$ so that $\sin \gamma \cos \sigma$ stays constant with $Z$, explicitly $\gamma \left( Z \right) = {\sin ^{ - 1}}\left[ {\sin \left( {\gamma \left( 0 \right)} \right)/\cos \left( {\sigma \left( Z \right)} \right)} \right]$, assuming that $\sigma(0)=0$. An example is shown in Figs. \ref{Fig:Figure4}(b-c) using a parabolic trajectory $f(Z)=0.03Z^2$ and $Z_m=40$. In (b) $\gamma = 25^0$ is kept fixed while in (c) it varies according to the previous formula. In (d) the FWHM of the two beams along $Z=ct.$ planes of the two beams are compared with the theoretical curves superposed. The desired effect of the increasing $\gamma$ in cancelling the increase of FWHM is evident.

We close by noting that an efficient experimental implementation of the proposed beams can be achieved along the lines of \cite{Froehly_2011}. In this context, the wave must first be generated in a large scale by reflecting a wide Gaussian beam on a spatial light modulator and the resulting beam is demagnified to the desired scale by means of  a strong reduction telescope.

\section{Conclusion}

The concept of Bessel-like accelerating beams, that was theoretically proposed \cite{Chremmos_2013_Bessel} and experimentally demonstrated \cite{Zhao_2013} with paraxial waves, has been extended to the nonparaxial regime. Following the key idea of \cite{Chremmos_2013_Bessel}, fully-vectorial nonparaxial accelerating Bessel-like waves were obtained by an appropriate modification of the conical ray pattern of standard Bessel beams. By applying the right phase mask to an unmodulated input wavefront, the "deformed" ray cones can be made to converge along a fairly arbitrary continuous focal line which defines the trajectory of the beam's main lobe. The phase required to produce a beam with a given trajectory and transverse width is calculated through a systematic procedure, that employs ray optics in the context of Rayleigh-Sommerfeld diffraction for the vector potential that generates the beam's EM components. Through the same ray optics approach, the distribution of the potential around the focus was shown to behave like a Bessel function, leading to similar distributions for the transverse ("strong") EM field components. When observed on a plane of constant $Z$, this distribution appears to be elliptically deformed along the direction of acceleration ($x$ in this paper), which is a result of  the fact that the ray cones are tilted at nonparaxial angles with respect to the $Z$ axis. Indeed, when the beam is observed on a plane that is normal to its trajectory at the current focal point, the standard circular Bessel pattern is obtained and it is of constant width, as long as the angle $(\gamma)$ of the ray cones has been selected to be $Z$-independent. The theoretical predictions were successfully verified through numerical simulations and a variety of different trajectories.

The beams introduced in this work can be useful in all applications mediated by the curved optical beams known so far, carrying the additional flexibility of \textit{arbitrary}, \textit{paraxial} and \textit{3D} trajectories. Moreover, due to their conical-interference structure, the new beams provide the convenience of a symmetric Bessel-like power profile which may be desired in applications and sets them clearly apart from optical-caustic beams, such as Airy beams and the like.

\section{Acknowledgement}
The work of I. D. Chremmos was supported by the project “Archimedes Center for Modeling, Analysis and Computation” (ACMAC, FP7-REGPOT-2009-1).
The work of N. K. Efremidis was supported by the action “ARISTEIA” in the context of the Operational Programme “Education and Lifelong Learning” that is co-funded by the European Social Fund and National Resources.

\bibliographystyle{ieeetr}
\bibliography{MyPapers}

\begin{thebibliography}{10}

\bibitem{Berry1979}
M.~Berry and N.~Balazs, ``{Non-spreading wavepackets},'' {\em Am. J. Phys.},
  vol.~47, no.~3, pp.~264--267, 1979.

\bibitem{Siviloglou2007}
G.~A. Siviloglou and D.~N. Christodoulides, ``Accelerating finite energy airy
  beams,'' {\em Opt. Lett.}, vol.~32, pp.~979--981, Apr 2007.

\bibitem{sivil-prl2007}
G.~A. Siviloglou, J.~Broky, A.~Dogariu, and D.~N. Christodoulides,
  ``Observation of accelerating {A}iry beams,'' {\em Phys. Rev. Lett.},
  vol.~99, p.~213901, Nov 2007.

\bibitem{Bandres2009}
M.~A. Bandres, ``Accelerating beams,'' {\em Opt. Lett.}, vol.~34,
  pp.~3791--3793, Dec 2009.

\bibitem{Unnikrishnan1996}
K.~Unnikrishnan and A.~R.~P. Rau, ``Uniqueness of the airy packet in quantum
  mechanics,'' {\em Am. J. Phys.}, vol.~64, no.~8, pp.~1034--1035, 1996.

\bibitem{Christodoulides_2008_Riding}
D.~Christodoulides, ``Optical trapping: Riding along an airy beam,'' {\em
  Nature Photonics}, vol.~2, no.~11, pp.~652--653, 2008.

\bibitem{Baumgart_2008}
J.~Baumgartl, M.~Mazilu, and K.~Dholakia, ``Optically mediated particle
  clearing using airy wavepackets,'' {\em Nature Photonics}, vol.~2, no.~11,
  pp.~675--678, 2008.

\bibitem{Zhang_2011_Morphing}
P.~Zhang, J.~Prakash, Z.~Zhang, M.~Mills, N.~Efremidis, D.~Christodoulides, and
  Z.~Chen, ``Trapping and guiding microparticles with morphing autofocusing
  airy beams,'' {\em Optics Letters}, vol.~36, no.~15, pp.~2883--2885, 2011.

\bibitem{Polynkin_2009}
P.~Polynkin, M.~Kolesik, J.~Moloney, G.~Siviloglou, and D.~Christodoulides,
  ``Curved plasma channel generation using ultraintense airy beams,'' {\em
  Science}, vol.~324, no.~5924, pp.~229--232, 2009.

\bibitem{Mathis_2012}
A.~Mathis, F.~Courvoisier, L.~Froehly, L.~Furfaro, M.~Jacquot, P.~Lacourt, and
  J.~Dudley, ``Micromachining along a curve: Femtosecond laser micromachining
  of curved profiles in diamond and silicon using accelerating beams,'' {\em
  Applied Physics Letters}, vol.~101, no.~7, 2012.

\bibitem{Salandrino_2010}
A.~Salandrino and D.~Christodoulides, ``Airy plasmon: A nondiffracting surface
  wave,'' {\em Optics Letters}, vol.~35, no.~12, pp.~2082--2084, 2010.

\bibitem{Efremidis2010}
N.~K. Efremidis and D.~N. Christodoulides, ``Abruptly autofocusing waves,''
  {\em Opt. Lett.}, vol.~35, pp.~4045--4047, Dec 2010.

\bibitem{Hu2012Springer}
Y.~Hu, G.~Siviloglou, P.~Zhang, N.~Efremidis, D.~Christodoulides, and Z.~Chen,
  {\em Self-accelerating Airy Beams: Generation, Control, and Applications},
  vol.~170 of {\em Springer Series in Optical Sciences}, pp.~1--46.
\newblock Springer, 2012.

\bibitem{Nye1999}
J.~F. Nye, {\em Natural Focusing and Fine Structure of Light: Caustics and Wave
  Dislocataions}.
\newblock IOP Publishing Ltd, 1999.

\bibitem{Greenfield2011}
E.~Greenfield, M.~Segev, W.~Walasik, and O.~Raz, ``Accelerating light beams
  along arbitrary convex trajectories,'' {\em Phys. Rev. Lett.}, vol.~106,
  p.~213902, May 2011.

\bibitem{Chremmos_2011_AAF}
I.~Chremmos, N.~K. Efremidis, and D.~N. Christodoulides, ``Pre-engineered
  abruptly autofocusing beams,'' {\em Opt. Lett.}, vol.~36, pp.~1890--1892, May
  2011.

\bibitem{Froehly_2011}
L.~Froehly, F.~Courvoisier, A.~Mathis, M.~Jacquot, L.~Furfaro, R.~Giust, P.~A.
  Lacourt, and J.~M. Dudley, ``Arbitrary accelerating micron-scale caustic
  beams in two and three dimensions,'' {\em Opt. Express}, vol.~19,
  pp.~16455--16465, Aug 2011.

\bibitem{Chremmos2012_PRA}
I.~D. Chremmos, Z.~Chen, D.~N. Christodoulides, and N.~K. Efremidis, ``Abruptly
  autofocusing and autodefocusing optical beams with arbitrary caustics,'' {\em
  Phys. Rev. A}, vol.~85, p.~023828, Feb 2012.

\bibitem{Chremmos_2011_Fourier}
I.~Chremmos, P.~Zhang, J.~Prakash, N.~Efremidis, D.~Christodoulides, and
  Z.~Chen, ``Fourier-space generation of abruptly autofocusing beams and
  optical bottle beams,'' {\em Optics Letters}, vol.~36, no.~18,
  pp.~3675--3677, 2011.

\bibitem{Piestun_1998}
R.~Piestun and J.~Shamir, ``Generalized propagation-invariant wave fields,''
  {\em J. Opt. Soc. Am. A}, vol.~15, pp.~3039--3044, Dec 1998.

\bibitem{Durnin1987}
J.~Durnin, ``Exact solutions for nondiffracting beams. i. the scalar theory,''
  {\em J. Opt. Soc. Am. A}, vol.~4, pp.~651--654, Apr 1987.

\bibitem{Mcgloin_2005}
D.~Mcgloin and K.~Dholakia, ``Bessel beams: Diffraction in a new light,'' {\em
  Contemporary Physics}, vol.~46, no.~1, pp.~15--28, 2005.

\bibitem{Chremmos_2013_Bessel}
I.~D. Chremmos, Z.~Chen, D.~N. Christodoulides, and N.~K. Efremidis,
  ``Bessel-like optical beams with arbitrary trajectories,'' {\em Opt. Lett.},
  vol.~37, pp.~5003--5005, Dec 2012.

\bibitem{Zhao_2013}
J.~Zhao, P.~Zhang, D.~Deng, J.~Liu, Y.~Gao, I.~D. Chremmos, N.~K. Efremidis,
  D.~N. Christodoulides, and Z.~Chen, ``Observation of self-accelerating
  bessel-like optical beams along arbitrary trajectories,'' {\em Opt. Lett.},
  vol.~38, pp.~498--500, Feb 2013.

\bibitem{Jarutis_2009}
V.~Jarutis, A.~Matijo\v{s}ius, P.~D. Trapani, and A.~Piskarskas, ``Spiraling
  zero-order bessel beam,'' {\em Opt. Lett.}, vol.~34, pp.~2129--2131, Jul
  2009.

\bibitem{Matijosius_2010}
A.~Matijo\v{s}ius, V.~Jarutis, and A.~Piskarskas, ``Generation and control of
  the spiraling zero-order bessel beam,'' {\em Opt. Express}, vol.~18,
  pp.~8767--8771, Apr 2010.

\bibitem{Morris_2010}
J.~Morris, T.~Cizmar, H.~Dalgarno, R.~Marchington, F.~Gunn-Moore, and
  K.~Dholakia, ``Realization of curved bessel beams: propagation around
  obstructions,'' {\em Journal of Optics}, vol.~12, no.~12, 2010.

\bibitem{Rosen_1995}
J.~Rosen and A.~Yariv, ``Snake beam: a paraxial arbitrary focal line,'' {\em
  Opt. Lett.}, vol.~20, pp.~2042--2044, Oct 1995.

\bibitem{Paterson_1996_Helicon}
C.~Paterson and R.~Smith, ``Helicon waves: propagation-invariant waves in a
  rotating coordinate system,'' {\em Optics Communications}, vol.~124,
  no.~1–2, pp.~131 -- 140, 1996.

\bibitem{Alonzo_2005_Helico}
C.~Alonzo, P.~J. Rodrigo, and J.~Gl\"{u}ckstad, ``Helico-conical optical beams:
  a product of helical and conical phase fronts,'' {\em Opt. Express}, vol.~13,
  pp.~1749--1760, Mar 2005.

\bibitem{Novitsky_2009}
A.~V. Novitsky and D.~V. Novitsky, ``Nonparaxial airy beams: role of evanescent
  waves,'' {\em Opt. Lett.}, vol.~34, pp.~3430--3432, Nov 2009.

\bibitem{Carretero_2009}
L.~Carretero, P.~Acebal, S.~Blaya, C.~Garc\'{i}a, A.~Fimia, R.~Madrigal, and
  A.~Murciano, ``Nonparaxial diffraction analysis of airy and sairy beams,''
  {\em Opt. Express}, vol.~17, pp.~22432--22441, Dec 2009.

\bibitem{Torre_2010}
A.~Torre, ``Airy beams beyond the paraxial approximation,'' {\em Optics
  Communications}, vol.~283, no.~21, pp.~4146 -- 4165, 2010.

\bibitem{Kaganovsky_2012}
Y.~Kaganovsky and E.~Heyman, ``Nonparaxial wave analysis of three-dimensional
  airy beams,'' {\em J. Opt. Soc. Am. A}, vol.~29, pp.~671--688, May 2012.

\bibitem{Courvoisier_2012_Femto}
F.~Courvoisier, A.~Mathis, L.~Froehly, R.~Giust, L.~Furfaro, P.~A. Lacourt,
  M.~Jacquot, and J.~M. Dudley, ``Sending femtosecond pulses in circles: highly
  nonparaxial accelerating beams,'' {\em Opt. Lett.}, vol.~37, pp.~1736--1738,
  May 2012.

\bibitem{Kaminer_2012}
I.~Kaminer, R.~Bekenstein, J.~Nemirovsky, and M.~Segev, ``Nondiffracting
  accelerating wave packets of maxwell's equations,'' {\em Phys. Rev. Lett.},
  vol.~108, p.~163901, Apr 2012.

\bibitem{Zhang_2012_Nonparaxial}
P.~Zhang, Y.~Hu, D.~Cannan, A.~Salandrino, T.~Li, R.~Morandotti, X.~Zhang, and
  Z.~Chen, ``Generation of linear and nonlinear nonparaxial accelerating
  beams,'' {\em Opt. Lett.}, vol.~37, pp.~2820--2822, Jul 2012.

\bibitem{Zhang_2012_Weber}
P.~Zhang, Y.~Hu, T.~Li, D.~Cannan, X.~Yin, R.~Morandotti, Z.~Chen, and
  X.~Zhang, ``Nonparaxial {M}athieu and {W}eber accelerating beams,'' {\em
  Phys. Rev. Lett.}, vol.~109, p.~193901, Nov 2012.

\bibitem{Aleahmad_2012_PRL}
P.~Aleahmad, M.-A. Miri, M.~S. Mills, I.~Kaminer, M.~Segev, and D.~N.
  Christodoulides, ``Fully vectorial accelerating diffraction-free {H}elmholtz
  beams,'' {\em Phys. Rev. Lett.}, vol.~109, p.~203902, Nov 2012.

\bibitem{Lax_1975}
M.~Lax, W.~H. Louisell, and W.~B. McKnight, ``From maxwell to paraxial wave
  optics,'' {\em Phys. Rev. A}, vol.~11, pp.~1365--1370, Apr 1975.

\bibitem{Balanis_2012}
C.~Balanis, {\em Advanced Engineering Electromagnetics}.
\newblock CourseSmart Series, Wiley, 2012.

\bibitem{GoodmanFourier}
J.~Goodman, {\em Introduction To Fourier Optics}.
\newblock McGraw-Hill physical and quantum electronics series, Roberts \&
  Company Publishers, 2005.

\end{thebibliography}

\end{document}